\documentclass{elsart}

\usepackage{natbib}

 \usepackage{epsfig}

\usepackage{amssymb}

\begin{document}

\begin{frontmatter}



\title{Direct Dark Matter Searches with CDMS and XENON\thanksref{footnote1}}
\thanks[footnote1]{Presented at the 36$\rm^{th}$ COSPAR Scientific Assembly in Beijing, July 2006}


\author{Kaixuan Ni\corauthref{cor}\thanksref{footnote2}}
\address{Department of Physics, Yale University, New Haven, CT, 06520, USA}
\corauth[cor]{Corresponding author}
\thanks[footnote2]{On behalf of the XENON collaboration}
\ead{kaixuan.ni@yale.edu}

\author{Laura Baudis\thanksref{footnote3}}
\address{Department of Physics, RWTH Aachen University, Aachen, 52074, Germany}
\thanks[footnote3]{On behalf of the CDMS and XENON collaborations}
\ead{laura.baudis@rwth-aachen.de}

\begin{abstract}

The Cryogenic Dark Matter Search (CDMS) and XENON experiments aim to directly detect dark matter in the form of weakly interacting massive particles (WIMPs) via their elastic scattering on the target nuclei. The experiments use different techniques to suppress background event rates to the minimum, and at the same time, to achieve a high WIMP detection rate. The operation of cryogenic Ge and Si crystals of the CDMS-II experiment in the Soudan mine yielded the most stringent spin-independent WIMP-nucleon cross-section ($\rm\sim10^{-43}~cm^2$) at a WIMP mass of 60\,GeV/$c^2$. The two-phase xenon detector of the XENON10 experiment is currently taking data in the Gran Sasso underground lab and promising preliminary results were recently reported. Both experiments are expected to increase their WIMP sensitivity by a one order of magnitude in the scheduled science runs for 2007.

\end{abstract}


\end{frontmatter}

\section{Introduction}

Recent observations from high-redshift supernovae \citep{Perlmutter:1998np, Knop:2003iy, Astier:2005qq}, cosmic microwave background anisotropy measurements \citep{Spergel:2003cb, Spergel:2006hy}, the red-shift galaxy clusters \citep{Peacock:2001gs}, and the Sloan Digital Sky Survey \citep{Tegmark:2003ud} provide growing evidence that the mass in the Universe is dominated by dark matter, which is non-luminous, non-baryonic and could be weakly interacting massive particles (WIMPs) \citep{Lee:1977}. One attractive candidate for WIMPs is the lightest supersymmetric particle (LSP) from supersymmetry models \citep{Jungman:1995df}. A WIMP can deposit a small amount of energy (a few tens of keV) in an elastic scattering with an atomic nucleus \citep{Goodman:1984dc}, with a rate less than 1 event/kg/day. 

Figure 1 shows the predicted event rates for a WIMP mass of 100 GeV/c$^2$ and a spin-independent WIMP-nucleon cross-section of $\rm10^{-43}~cm^2$ for different target materials, according to \citet{Lewin:1995rx}. It can be seen that the event rate falls quickly at high recoil energy. A low energy-threshold ($\sim$10~keV) is one of the key requirements for direct dark matter search experiments, such as CDMS and XENON, to achieve a good WIMP detectability. While the WIMP detection rate on xenon (the target material for XENON experiment) is suppressed at large recoil energies (we use the nuclear form factor by \citet{Engel:1991wq}), it is enhanced at low recoil energies by the large atomic mass of xenon. From figure 1, the event rate in a Xe detector is about 30\% higher than the one in Ge (one of the two target materials for CDMS experiment, another one is Si), for an energy-threshold of 10 keV.

\section{CDMS-II Experiment}
The CDMS-II experiment (a detailed description of the experimental apparatus can be found in \citet{Akerib:2005zy}) is located in one of the two excavated caverns of the Soudan Underground Laboratory, at a depth of 780 m (2090 m.w.e. or meter-water-equivalent). The surface muon flux is reduced at the underground lab by a factor of 5 $\times10^4$. An additional active muon veto shield, consisting of 5-cm-thick BC-408 plastic scintillator slabs, is used to tag cosmic ray muons. A passive shield, consisting of 40-cm-thick outer polyethylene, 22.5-cm-thick lead (including 4.5-cm-thick inner ancient lead), and 10-cm-thick inner polyethylene, is enclosed by the muon veto and houses the detectors. The passive shield reduces most of the gamma-ray and neutron background from radioactivity in the surrounding rock or produced by cosmic ray muons interacting with the surrounding rock and shield. To reduce the radon activity in the vicinity of the detectors, medical grade breathing air is used to purge the space in the shield continuously.

At the core are Z(depth)-sensitive ionization and phonon-mediated (ZIP) detectors, kept at a base temperature of about 50 mK. Each ZIP detector is a cylindrical high-purity Ge (250 g) or Si (100 g) crystal. Two concentric ionization electrodes and four independent phonon sensors are photolithographically patterned onto each crystal. A particle interacting in a ZIP detector causes either an electron recoil (gamma rays, electrons, neutron inelastic scattering, etc.) or a nuclear recoil (neutron elastic scattering, WIMPs). The interaction deposits its energy into the crystal through charge excitations (electron-hole pairs) and lattice vibrations (phonons). The charge excitations are collected with electrodes on the two sides of the ZIP detector. The electron-equivalent energy $E_Q$ is equal to $N_Q \times \epsilon$, where $N_Q$ is the number of collected charges and $\epsilon$ is the average energy needed to produce an electron-ion pair ($\epsilon \approx 3$ eV in Ge and 3.8 eV in Si). The phonon signal is detected by electrothermal-feedback transition-edge sensors (QETs) photolithographically patterned onto one of the crystal faces. Since the drifting of electrons and holes across the crystal also contributes the phonon signal, the recoil energy $E_R$ of an event is inferred from both the phonon and ionization signals. $E_Q$ is equal to the recoil energy $E_R$ for electron recoils. Nuclear recoils produce less electron-ion pairs, which makes $E_Q$ less than $E_R$. Thus the ionization yield, defined as $y = E_Q/E_R$, is much smaller for nuclear recoils ($y\sim0.3$ for Ge and $\sim0.25$ for Si above 20 keV) than that for electron recoils ($y = 1$). It provides the technique to reject the electron-recoil events from most of the background.

The CDMS-II experiment has been operating in the Soudan mine since October 2003 and reported their recent results, from WIMP search data collected through August 2004 with a total of 74.5 live days exposure, yielding a total spectrum-weighted exposures of 34 (12) kg d for the Ge (Si) targets in the 10-100 keV nuclear recoil region \citep{Akerib:2005kh}. Analysis was performed blindly by masking the events from the candidate region, defined by using calibration data from $^{133}$Ba and $^{252}$Cf sources and from non-masked WIMP search data. The ZIP detector provides excellent event-by-event discrimination of nuclear recoils from background electron recoils in the detector's bulk region. However, electron recoils near the detector's surface suffer from poor ionization collection and leak into the candidate nuclear recoil region. New analysis techniques were developed to remove those surface electron recoils. The phonon pulses from surface recoils are more prompt than those recoils in the detector bulk. A timing parameter, defined as the sum of the time delay of the phonon signal relative to the fast ionization signal and the phonon rise time, was used to reject the surface electron recoils. One candidate event was found after unmasking the Ge WIMP search data (Figure 2 top). However, that event occurred in a time period when the detector suffered inefficient ionization collection. It is also consistent with the rate of expected background. The expected number of surface events after the timing cut is 0.4$\pm$0.2(stat)$\pm$0.2(sys) between 10-100 keV in Ge. The expected neutron background that escapes the muon veto is 0.06 events in Ge. No candidate event was found in the Si data (Figure 2 bottom). Based on these data and previous results from Soudan \citep{Akerib:2004fq}, a 90\% C.L. upper limit on the spin-independent WIMP-nucleon cross section is $\rm1.6\times10^{-43}~cm^2$ from Ge and $\rm3\times10^{-42}~cm^2$ from Si, for a WIMP mass of 60 GeV/$c^2$.

The experiment is now operating five towers (21 Ge and 9 Si ZIP detectors) with more than 4 kg of Ge and 1 kg of Si and is expected to reach a sensitivity down to $\rm2.1\times10^{-44}~cm^2$ for spin-independent WIMP-nucleon cross section at 60 GeV/$c^2$ in 2007 \citep{Cabrera:2006}.

\section{XENON10 Experiment}
The XENON experiment uses two-phase (liquid/gas) xenon detectors to search for WIMPs \citep{Aprile:2002ef}. In the core of the detector, there is a bulk of liquid xenon (LXe) as the target for WIMP interactions. The relatively low costs of xenon and the flexibility of condensing gas to liquid phase make it feasible to build a large-mass detector. 

The electron/nuclear recoil from a particle interaction produces excitation ($Xe^*$) and ionization (electron-ion pair $Xe^++e^-$) in the liquid. The excited Xe molecules decay and produce scintillation light in the UV range (175 nm). The XENON experiment uses UV-sensitive photo-multipliers (PMTs) to detect the prompt light signal (also called primary light, or $S1$). The electron-ion pairs recombine and form excitation states, which also contribute to $S1$. By applying a drift field ($\sim$ 1 kV/cm) in LXe, part of the ionization electrons can be liberated from recombination. Those electrons are drifted to the gas phase, where they produce secondary excitation with a gain of a few hundred \citep{Bolozdynya:1999} in a strong electric field ($\sim$10 kV/cm). The scintillation light from the secondary excitation in the gas is also detected by the PMTs as a delayed signal (also called secondary light, or $S2$). The two-phase operation allows detection of a single ionization electron from one event. The design of the XENON10 detector also allows an efficient $S1$ collection ($\sim$ 1 photoelectron/keVr), leading to a conservative nuclear recoils threshold of 10~keVr.

Due to the difference in stopping power and ionization density of electron and nuclear recoils in LXe, the ratio of ionizing electrons escaping the electron-ion recombination is different, resulting in smaller $S2/S1$ values for nuclear recoils (Figure 3). This is the key technique for nuclear/electron recoil discrimination in XENON. It has already been demonstrated in small two-phase prototypes, for which a background discrimination of $>$ 98.5\% down to 20 keVr has been achieved \citep{Aprile:2006kx}. Further investigations indicate that an electron recoil discrimination of more than 99\% can be obtained with a 50\% nuclear recoil acceptance down to 10 keVr \citep{Shutt:2006ed}.

The XENON10 detector, with a target LXe mass of 15 kg, is currently operating in Gran Sasso Underground Laboratory, which is off a high-way tunnel under 1400 meter rock overburden (3500 m.w.e.). The total muon flux is about 1 /m$^2$/hr, one-order of magnitude lower than that in the Soudan mine. The detector is installed in a passive shield consisting a 20-cm-thick outer layer of Pb (including 5 cm of low-activity Pb with 30 Bq/kg of $^{210}$Pb) and 20-cm-thick inner layer of polyethylene, reducing the external gamma flux by a factor of $10^5$ and the neutron flux by a factor of 100 \citep{Sorensen:idm06}. Radon purge from N$_2$ gas is used in the shield continuously. The background rate from $^{85}$Kr contamination (25 ppm Kr in natural Xe) is reduced by a factor of 5000 by using a commercial available low-Kr (5 ppb) xenon, contributing negligible rate compared to the background rates from the detector materials, such as the PMTs and stainless steel vessels. However, the XENON10 detector provides 3D position information \citep{Ni:2006rj} and the self-shielding of LXe reduces the background rate in the central region of the LXe target by more than one order of magnitude for energies below 50 keVee (electron-equivalent energy). Preliminary Monte Carlo simulations predict less than one leakage event in 5 to 15 keVee (corresponding to roughly 15 to 40 keVr) in every 50 kg d exposure of the XENON10 detector. 

The collaboration recently reported a preliminary result with no candidate event in 18 kg d exposure down to 15 keVr nuclear recoil energy \citep{Yamashita:idm06}. The continuous stable operation of the detector for several months will allow XENON10 to reach a WIMP detection sensitivity similar to the current CDMS-II limit. Further upgrades, including replacement of several radioactive ``hot" components in the detector, will allow XENON10 to explore the WIMP-nucleon cross section down to the $\rm10^{-44}~cm^2$ region in 2007.

\section{Summary}
The development of low radioactive, low energy-threshold, and high background discrimination detectors by the CDMS and XENON collaborations provides promising ways to explore new physics by searching for dark matter particles via their direct elastic scattering in terrestrial targets. The recent CDMS result starts to exclude parameter space from supersymmetry models, as shown in Figure 4. The XENON10 experiment is expected to report first results from the current science run in 2006. Both experiments should reach a factor of 10 increase in sensitivity with respect to the current CDMS level by the end of next year.





\clearpage

\begin{figure}
\label{fig1}
\begin{center}
\includegraphics*[width=10cm,angle=0]{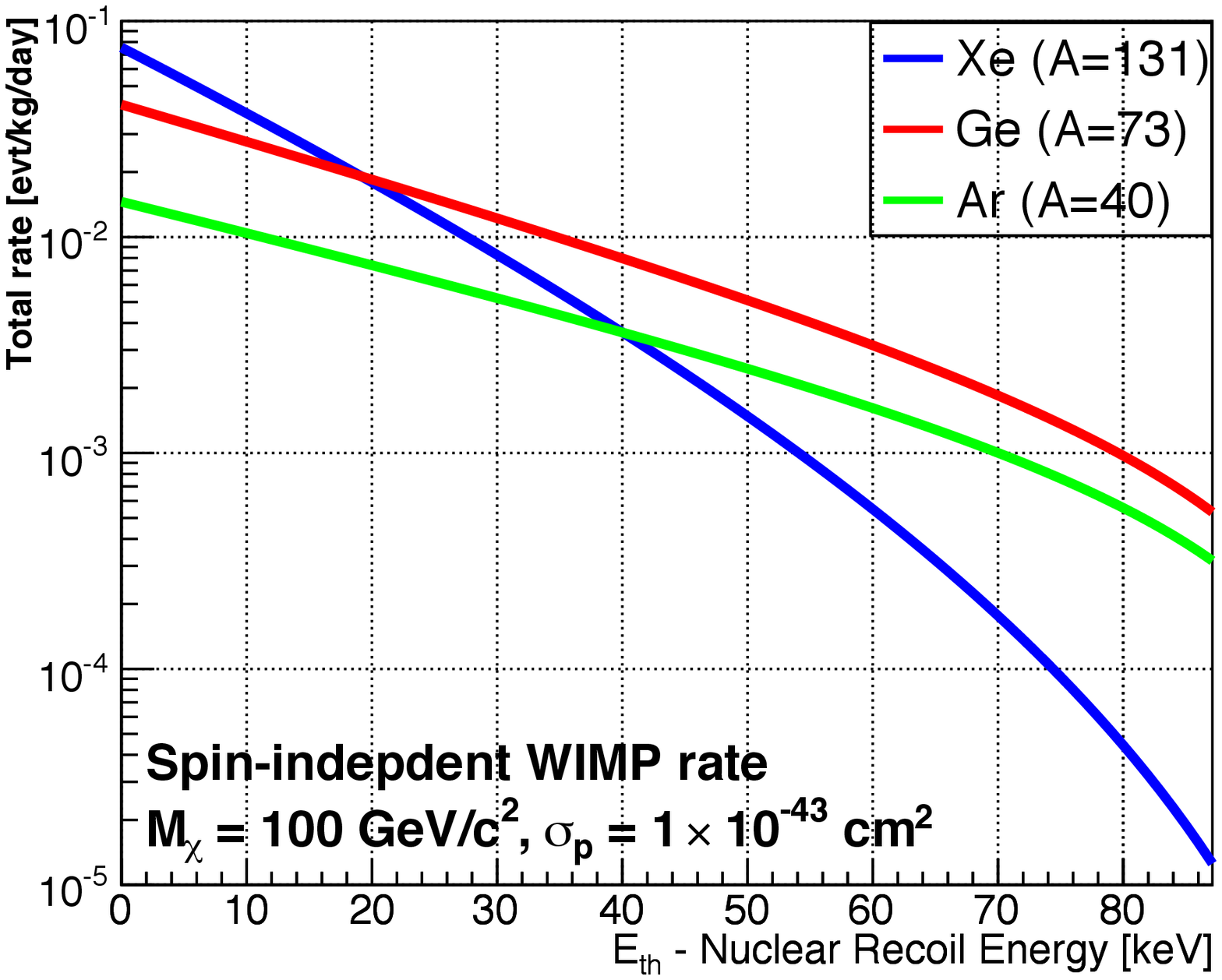}
\end{center}
\caption{Calculated WIMP detection rate as a function of detector energy-threshold $E_{th}$ for three different target materials (Xe, Ge and Ar). A WIMP mass of 100~GeV/$c^2$ and WIMP-nucleon cross-section of $\rm10^{-43}~cm^2$ are used.}
\end{figure}

\begin{figure}
\label{fig2}
\begin{center}
\includegraphics*[width=10cm,angle=0]{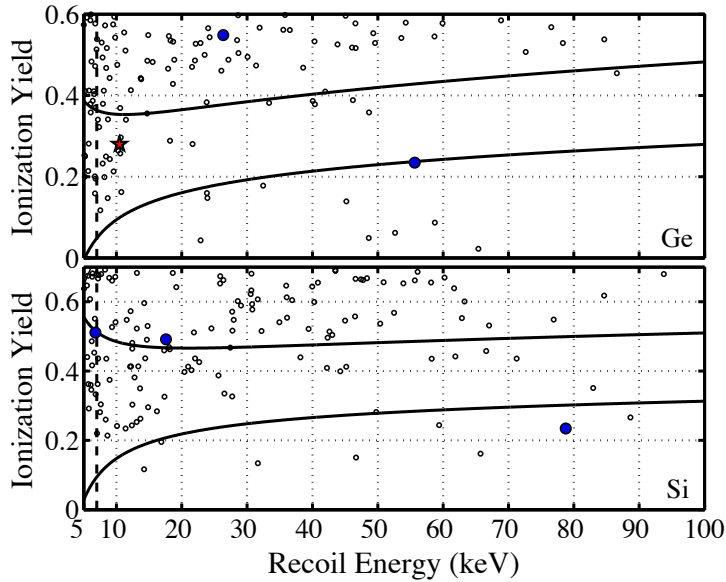}
\end{center}
\caption{Ionization yield versus recoil energy for events in the CDMS-II (top - Ge, bottom - Si) detectors, before the timing cut for removing the surface electron recoil events. The curved lines represent the defined WIMP signal regions with an energy-threshold at 7 keV (dashed vertical lines). After the cut, one event (star, red) remains in the signal region for the Ge data, while no candidate event is seen for the Si data. Figure from \citet{Akerib:2005kh}.}
\end{figure}

\begin{figure}
\label{fig3}
\begin{center}
\includegraphics*[width=10cm,angle=0]{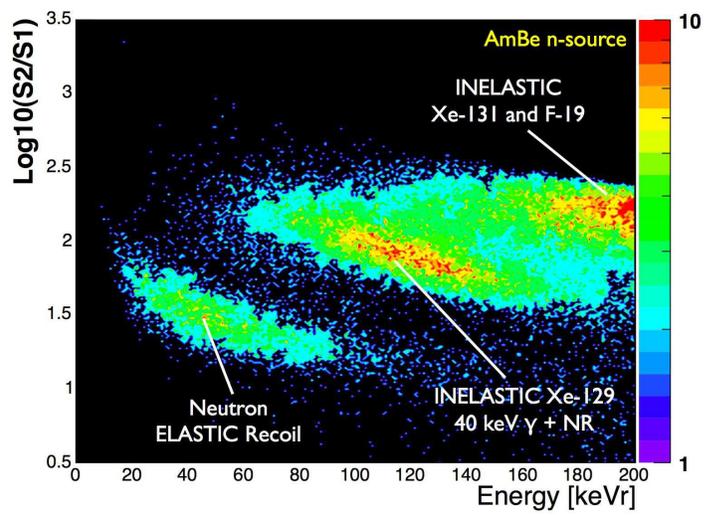}
\end{center}
\caption{Response of a two-phase Xe detector to an AmBe neutron source. The $S2/S1$ values from nuclear recoil events are clearly smaller than those from electron recoils, demonstrating the capability of two-phase Xe technique for background rejection. Figure from \citet{Aprile:2006kx}.}
\end{figure}

\begin{figure}
\label{fig4}
\begin{center}
\includegraphics*[width=10cm,angle=0]{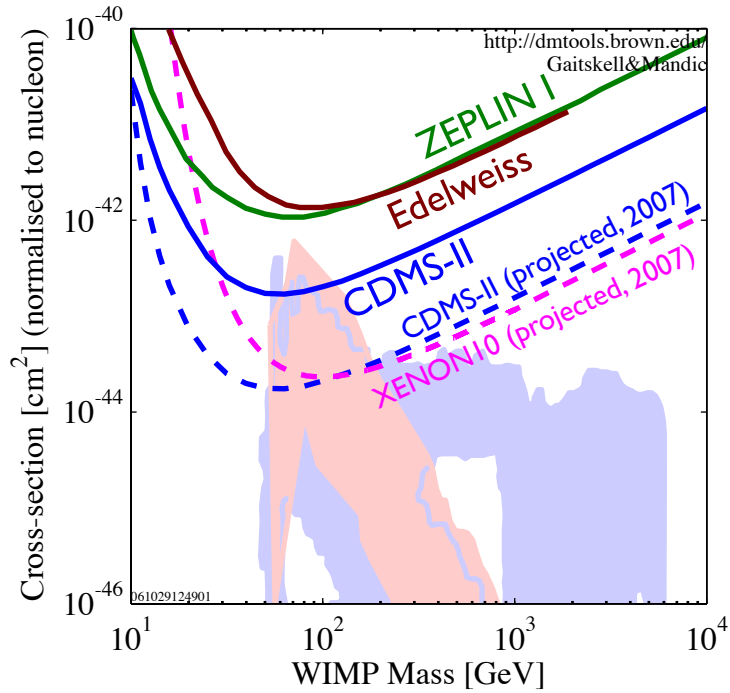}
\end{center}
\caption{WIMP-nucleon cross-section upper limit (90\% C.L.) from direct dark matter search experiments CDMS \citep{Akerib:2005kh}, Edelweiss \citep{Sanglard:2005we} and ZEPLIN I \citep{Alner:2005pa}. The projected upper limits from CDMS-II and XENON10 experiments in 2007 are shown as dashed lines. The shaded areas are allowed parameters from some supersymmetry models \citep{Baltz:2004aw, Ellis:2005mb}.}
\end{figure}

\end{document}